# Resistance Anomaly in Disordered Superconducting Films

J. Hua and Z. L. Xiao[*]

Materials Science Division, Argonne National Laboratory, Argonne, Illinois 60439
and Department of Physics, Northern Illinois University, DeKalb, Illinois 60115

D. Rosenmann[1], I. S. Beloborodov[12], U. Welp[1], W. K. Kwok[1], and G. W. Crabtree[1]

[1]Materials Science Division, Argonne National Laboratory, Argonne, Illinois 60439
[2]Department of Physics, University of Chicago, Chicago, Illinois 60637

We report on a resistance anomaly in disordered superconducting films containing arrays of irregularly distributed nanoscale holes. At high driving currents, peaks appear in the resistance as a function of temperature, with peak values up to 2% above the classic normal-state resistance. We attribute the observed resistance anomaly to dissipation-induced granularity which enhances the contributions from fluctuation-induced reduction of the density of states of the quasiparticles. The granular feature of a disordered superconducting film originates from the inhomogeneous temperature distribution caused by the variation of the local dissipation and/or heat transfer.

---

[*] zxiao@niu.edu or xiao@anl.gov

The physical properties of superconducting films can be modified by introducing hole-arrays into them. For example, superconducting films containing periodic hole-arrays (antidot arrays) can become wire networks which behave totally different from continuous films [1-2]. Regular hole-arrays have also been considered to be effective pinning centers to increase the critical current of a superconducting film [3-7]. On the other hand, disordered superconducting antidot arrays have hardly been explored in experiments. Here we present transport measurements on disordered superconducting antidot arrays achieved by utilizing substrates which contain networks of irregularly distributed nanoscale channels fabricated through self-assembly using an electrochemical process. We observed an intriguing resistance anomaly, which manifests itself as bumps or peaks in the resistance versus temperature ($R\sim T$) curves at high driving currents. The observed peak value can be larger than the classic normal-state resistance. Excess resistances have been reported in superconducting Al nanowires [8-10], nanoloops [10-11] and nanodiscs [12] where they were believed to originate from normal-superconducting ($N$-$S$) interfaces induced either by dynamic phase slip centers [8-10] or different critical temperatures ($T_c$) in the neighboring parts [10-12]. However, the resistance anomaly reported here in our disordered superconducting antidot arrays can best be understood with dissipation-induced granurality in which fluctuation-induced reduction of the density of state (DOS) of the quasiparticles is enhanced, resulting in the observed excess resistance.

Figure 1(a) shows a typical scanning electron microscopy (SEM) image of the studied disordered superconducting antidot arrays. The substrates were anodic aluminum oxide (AAO) membranes obtained by anodizing high purity aluminum foils in a selected acid [13]. Parallel nanoscale channels form in the aluminum oxide during the anodization. The size and distribution of the holes depend strongly on the applied anodization voltage and growth time. Detailed information on the general fabrication process of AAO membranes used in this research is reported elsewhere [13,14]. The AAO membrane (Anodisc 13) has an average hole diameter of 200 nm [14]. Superconducting films with irregular arrays of holes were fabricated by depositing niobium onto these AAO membranes. The base vacuum for sputtering Nb is better than $10^{-9}$ Torr and argon served as the working gas at a pressure of 1.5 mTorr. The thickness of the Nb films is 100 nm deposited at rate of 2 Å/s. The substrates were not heated during the deposition.

Four-probe DC transport measurements were carried out on rectangular shaped samples, 0.9 ~1.5 mm wide and 3 ~ 5 mm long. Attention was paid to ensure that the current and voltage leads were aligned on the same line [15]. The resistance measurements were conducted on three samples containing irregularly distributed nanoholes and on one sample with an ordered hole array (see Ref.7 for a SEM image) for comparison. Resistance anomalies were only observed in the disordered films.

Our main results are presented in Figs. 2 and 3: anomalous resistance bumps and peaks appear in the temperature dependent resistance curves at high driving currents, both in the absence and presence of an external applied field. It is known that in bulk and thin film Nb, the transition from normal to superconducting state is associated with a monotonic decrease of resistance with decreasing temperature. As can be seen in Fig.2, this behavior also holds at small currents in our disordered Nb superconducting antidot arrays. However, a small resistance bump appears when the driving current reaches a certain critical value ($I > 6.0$ mA). With increasing driving currents the resistance bump evolves into a peak whose value becomes larger than $R_N$, with $R_N$ being the normal-state resistance in the absence of fluctuations. In the explored temperature range ($> 4.2$ K), this excess resistance can increase up to 2% of $R_N$.

Although interpretations on the excess resistance observed in Al nanostructures are still qualitative and under debate, static or dynamic *N-S* interfaces has been considered to play a key role in all the proposed models [8-12]. In a disordered superconducting antidot array, $T_c$ variation can exist due to the spatial variation in Nb sections adjoining the irregularly spaced holes. The spatial variations could induce *N-S* interfaces resulting in the appearance of an excess resistance. For example, as shown in Fig.3(a), the amplitude of the resistance peaks decreases with increasing magnetic fields, similar to that observed in Al nanostructures [9]. However, all *N-S* interface based models predict a suppression of the anomalous resistance peak with increasing driving currents, leading to the disappearance of the peak at high currents [5,9]. In contrast, the excess resistance *increases* with current in our disordered superconducting antidot arrays. Hence the resistance anomaly observed here may originate from a different mechanism. A strong evidence for this supposition is the appearance of a second sharp resistance drop following the resistance peak at a lower temperature as shown in Fig.2(a) where complete *R-T* curves over a large temperature range are presented. According to the *N-S* interface mechanism, a resistance peak or bump would occur at the lower resistance step where the superconducting antidot array is partially superconducting and normal. Furthermore, the *N-S* interface based excess resistance can only be detected by placing superconducting voltage leads very close to the interface, since the effect originates from the potential difference of quasiparticles and Cooper pairs in the vicinity of the interface.

One salient feature shown in Fig. 2, is that the $R \sim T$ curves shift to lower temperatures with increasing driving current. This can be the sign of self-heating. The appearance of discontinuities/jumps and hysteresis in $R \sim T$ curves support this idea, since self-heating can induce bi-stability in superconductors [16]. In this case, at a bath temperature $T_{j1}$ (see Fig.3(b) for definition) which is measured by the thermometer, the actual temperature of the sample is $T_N$ which is the tangent point of the self-heating release $Q = j^2\rho$ and the heat transfer $W = h(T - T_{j1})/d^*$. Here, $j$ is the current density, $\rho$ the resistivity, $h$ is the heat transfer coefficient and $d^* = A/p$ is the effective thickness for heat transfer with $p$ and $A$ being the perimeter and area of the sample cross-section, respectively [16]. Once the temperature is below $T_{j1}$, $Q$ becomes smaller than $W$, leading to a sharp decrease in the resistance because the actual sample temperature drops from $T_N$ to a value equal or close to the bath temperature. Because $T_N > T_c$, the entire sample should be in the *normal state* at a bath temperature higher than $T_{j1}$. Thus, the observed resistance peaks, which appear in the normal state, cannot originate from the *N-S* interface or from dynamic phase slip mechanisms. However, superconducting fluctuations can decrease or increase the resistance at $T > T_c$ [17,18]. Below we show that superconducting fluctuations can account for the main features of the experimental data presented here.

The two sharp resistance drops imply that the sample may consist of two parts, distinguished by different local $T_c$ or disparate local sample temperatures for a fixed bath temperature. We designate part I and II as being associated with the first and second resistance jumps at $T_{j1}$ and $T_{j2}$, respectively. Since the absence of peaks at the lower resistance steps excludes the inhomogeneity of the critical temperature, the jumps at $T_{j1}$ and $T_{j2}$ can be attributed to the spatial inhomogeneity of the sample temperature caused by variation of local heat production and/or heat transfer. In the latter case, the local heat transfer $W$ depends on the width of the adjoining Nb sections between holes, because the contribution from the side edges to heat transfer is comparable to that from the top and bottom surfaces of the Nb sections, i.e. $d^*$ will depend on the width of that section. For example, $d^*$ will be 25 nm and 33 nm for 100 nm thick sections of 100 and 200 nm in width respectively, leading to a 30% better heat transfer for the narrower section. Since at $T_{j1}$ and $T_{j2}$ the sample temperature is equal to $T_N$ for both parts of the sample, the difference $\Delta T_j = T_{j1} - T_{j2}$ reflects the temperature of the two parts of the sample. As shown in the inset of Fig.3(b), $\Delta T_j$ increases with increasing driving current.

The transition to the superconducting state for part II (corresponding to the lower temperature resistance step) can be interpreted as a conductivity increase $\Delta\sigma_{AL}$ induced by non-equilibrium Cooper pairs as predicted by Aslamazov-Larkin [17,18]. The curve fitting with the AL contribution $R_{AL(II)} = R_{II}/[1+\lambda T_c/(T-T_c)]$ are shown as thin dashed line in Fig. 3(b), where $\lambda \sim (\tau T_c)^{1/2}$ with $\tau$ being the electron relaxation time and $\lambda << 1$. $R_{AL(II)}$ is the resistance with AL contribution for part II, and $R_{II}$ is the normal-state resistance for this part of the sample in the absence of fluctuations [17,18]. In the theoretical formula, $T$ represents the sample temperature, while in our curve fitting, $T$ is the measured bath temperature. Therefore, $T_c$ was replaced with $T_{j2}-\Delta T_0$, where $\Delta T_0 = T_N-T_c$ is a fitting parameter. That is, at a bath temperature $T_{j2}-\Delta T_0$ the sample is actually at the critical temperature $T_c$ due to self-heating. The extracted fitting parameters $\lambda$ and $\Delta T_0$ for the curves obtained in 0.5 T and various currents are given in the inset of Fig.3(b), where $R_{II} = 3.0 \ \Omega$ was used for all fittings.

At temperatures above $T_{j1}$, the temperature in part I is also above $T_c$. The resistance for this part with the AL contribution should follow the same temperature dependence as in part II, i.e. $R_{AL(I)} = R_I/[1+\lambda T_c/(T-T_c)]$. If there were no contributions from other mechanisms, the total resistance $R$ should have been simply equal to the sum of $R_{AL(I)}+R_{AL(II)}$, as demonstrated by the thick dashed curve in Fig. 3(b) for the $R$-$T$ curve obtained at 8.5 mA. In fitting the data for part I, $T_c$ was replaced with $T_{j1}-\Delta T_0$ and the same $\lambda$ and $\Delta T_0$ as those for part II were used. We obtained $R_I = 0.91 \ \Omega$. The curve which includes only the AL contribution deviates clearly from the measured one in the vicinity of the transition and, more importantly, it lacks the peak feature.

At $T > T_c$, a resistance with a value larger than $R_N$ can be induced by the DOS contribution while a peak can appear in the $R \sim T$ curve due to the competition between the AL and the DOS contributions. However, the absence of peaks for part II of the sample indicates that the DOS contribution is not large enough to overcome the AL contribution in a homogeneous superconductor. On the other hand, Beloborodov et al. [18] proposed to use granularity enhanced DOS contribution to understand the excess resistance observed in 3D granular superconductors where the granularity induced DOS contribution is much larger than the AL contribution. The peaks in our disordered superconducting films can also be understood with granularity enhanced DOS contributions: at $T > T_{j1}$, the temperature in part I of the sample is closer to $T_c$ than in part II which plays the role of the normal metal in the theoretical formalism. That is, the disordered superconducting film behaves like a granular superconductor at $T > T_{j1}$ due to the inhomogeneous temperature distribution. Furthermore, an increase of $\Delta T_j$ means that the temperature in part II of the sample is farther away from $T_c$, causing a weakening of the coupling strength between neighboring part I. Similar to what is predicted in a 3D granular superconductor [18], such a decrease of the coupling strength increases the DOS contribution. This implies that by adjusting the driving current one can change the DOS contribution by utilizing the temperature distribution in a disordered superconducting film. Hence, the DOS contribution increases with current due to the larger temperature differences between parts I and II and eventually surpasses the AL contribution, leading to a resistance peak in the transition. Experimentally, the resistance with DOS contribution shown in Fig.3(b) as a thick solid curve, can be derived with the measured $R$ and the calculated $R_{AL}$ (=$R_{AL(I)} + R_{AL(II)}$) and follows the expected temperature dependence [18].

In conclusion, we observed a resistance anomaly in disordered superconducting films and interpreted its origin as dissipation-induced granularity arising from the inhomogeneous heat-transfer and/or local dissipation in the sample. This provides not only an alternative mechanism to account for excess resistances but also a way to study superconducting fluctuations, especially those related to a sample's granularity, since the coupling strength in a disordered superconducting film can be conveniently adjusted with the driving current.

We thank F. M. Peeters and M. V. Milosevic for stimulating discussions. This work was supported by NSF Grant No. DMR-0605748 and by the US Department of Energy under award No. DE-FG02-06ER46334 and contract DE-AC02-06CH11357. SEM imaging was performed in the Electron Microscopy Center at Argonne.

**Figure captions**

Fig. 1. SEM image of a disordered superconducting antidot array.

Fig. 2. (color online) (a) $R \sim T$ curves at 0.5 T and various currents. (b) an expansion of the data near $R_N$ to show the details of the bumps/peaks. The inset of (b) shows typical hysteresis in a $R \sim T$ curve ($I = 8.5$ mA) with arrows indicating the temperature sweeping directions.

Fig. 3. (color online) (a) $R \sim T$ curves at 8 mA and various magnetic fields. (b) Analysis of the $R \sim T$ curve obtained at 8.5 mA and 0.5 T. Definitions of $T_{j1}$, $T_{j2}$ and $\Delta T_0$ are also given. Open circles are experimental data, thin dashed curve represents the fitting with the AL contribution for the lower resistance step. The derived $\Delta T_0$ (=$T_{j2}$-$T_c$) and $\lambda$ for 8.5 mA (and also for other currents) are given in the inset along with $\Delta T_j$ =$T_{j2}$-$T_{j1}$. The thick dashed curve in (b) is the calculated resistance with AL contributions for the whole sample at 8.5 mA based on the fitting data ($\lambda$ and $\Delta T_0$) for the lower resistance step. The thick solid curves represent the resistances with DOS contributions. $R_{II} = 3.0$ $\Omega$ and $R_I = 0.91$ $\Omega$ were used for all fittings. Please see text for details.

**FIG.1**

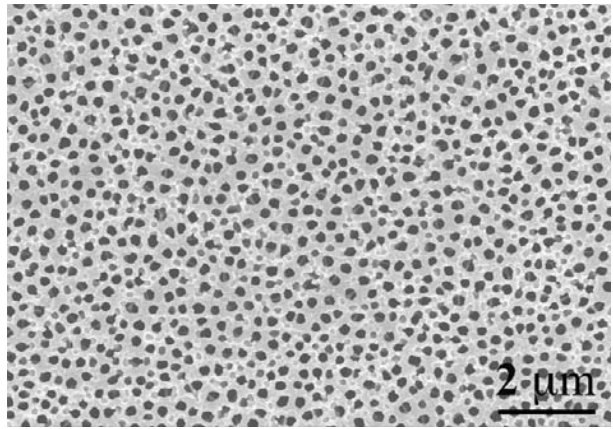

**FIG.2**

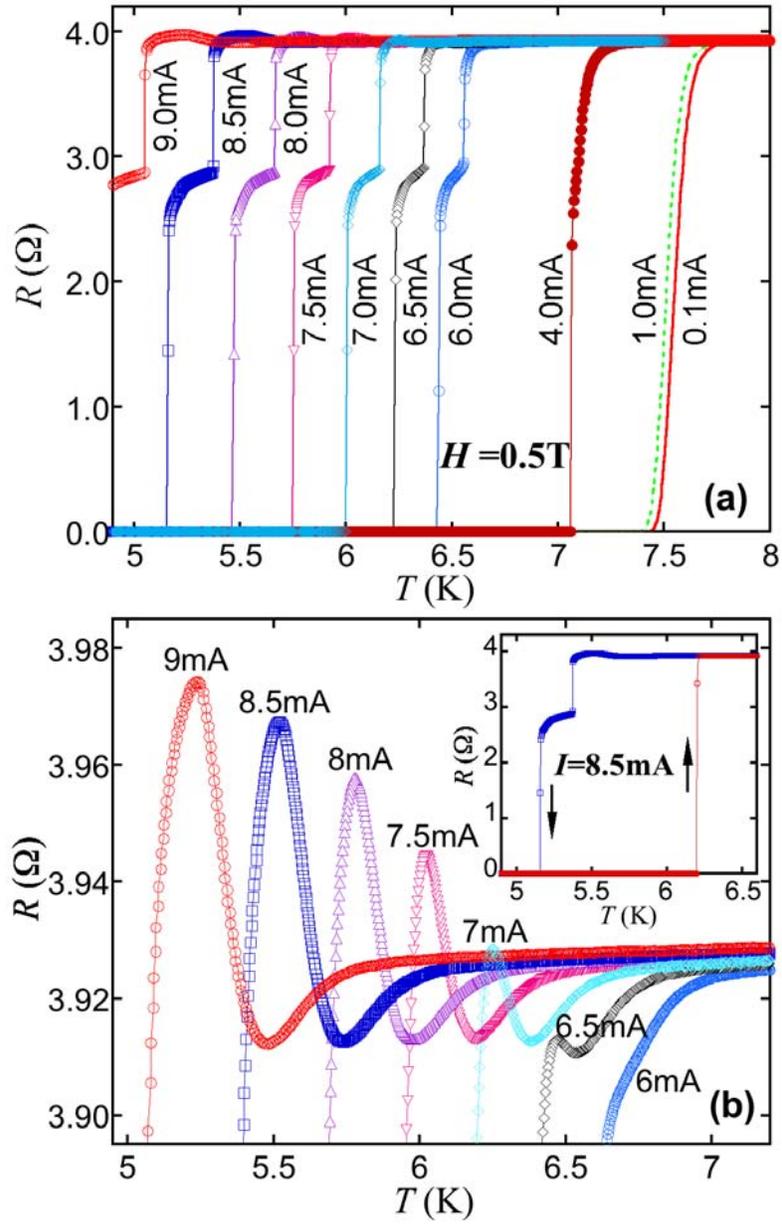

**FIG.3**

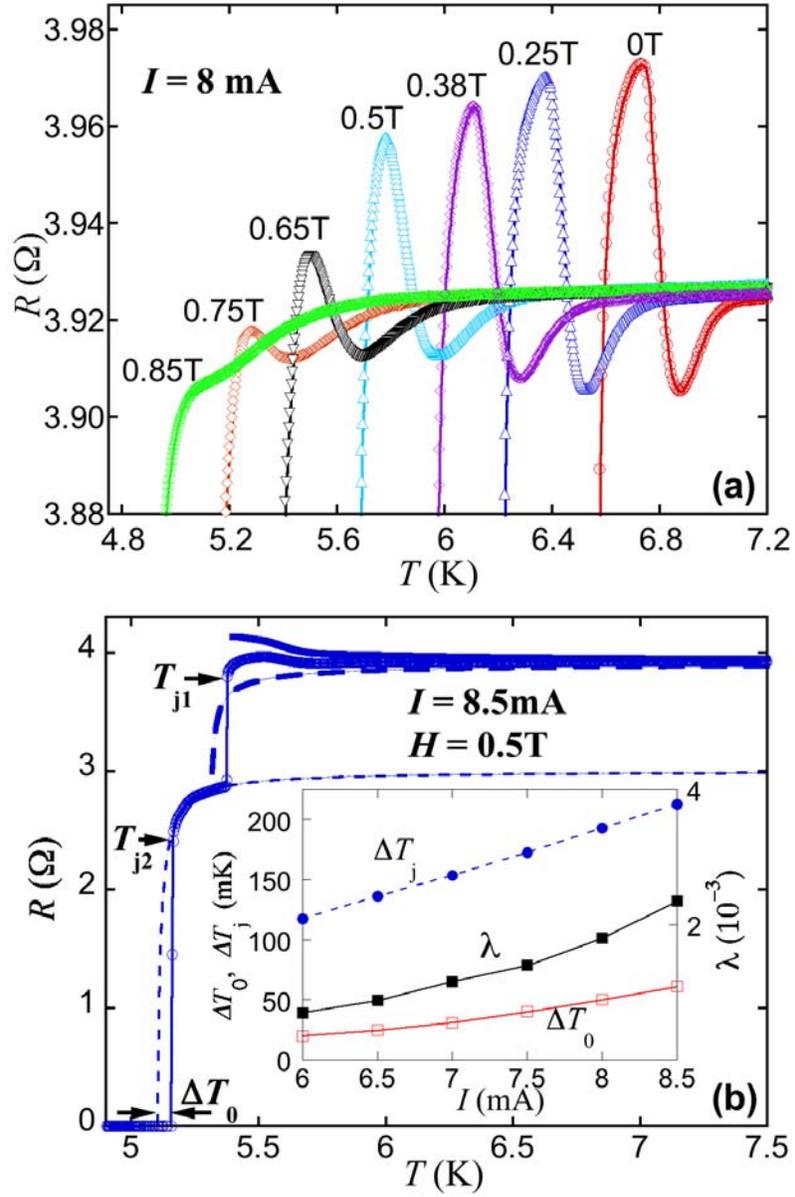